\documentstyle[aps,eqsecnum,preprint,floats,epsf,epsfig]{revtex}
\begin{document}
\def\be{\begin{eqnarray}}
\def\en{\end{eqnarray}}
\def\non{\nonumber}
\def\la{\langle}
\def\ra{\rangle}
\def\nc{N_c^{\rm eff}}
\def\vp{\varepsilon}
\def\a{{\cal A}}
\def\B{{\cal B}}
\def\c{{\cal C}}
\def\d{{\cal D}}
\def\e{{\cal E}}
\def\p{{\cal P}}
\def\t{{\cal T}}
\def\up{\uparrow}
\def\dw{\downarrow}
\def\vma{{_{V-A}}}
\def\vpa{{_{V+A}}}
\def\smp{{_{S-P}}}
\def\spp{{_{S+P}}}
\def\J{{J/\psi}}
\def\ov{\overline}
\def\Lqcd{{\Lambda_{\rm QCD}}}
\def\pr{{\sl Phys. Rev.}~}
\def\prl{{\sl Phys. Rev. Lett.}~}
\def\pl{{\sl Phys. Lett.}~}
\def\np{{\sl Nucl. Phys.}~}
\def\zp{{\sl Z. Phys.}~}
\def\lsim{ {\ \lower-1.2pt\vbox{\hbox{\rlap{$<$}\lower5pt\vbox{\hbox{$\sim$}
}}}\ } }
\def\gsim{ {\ \lower-1.2pt\vbox{\hbox{\rlap{$>$}\lower5pt\vbox{\hbox{$\sim$}
}}}\ } }

\font\el=cmbx10 scaled \magstep2{\obeylines\hfill March, 2003}

\vskip 1.5 cm

\centerline{\large\bf Hadronic Charmed Meson Decays}
\centerline{\large\bf Involving Tensor Mesons}
\bigskip
\centerline{\bf Hai-Yang Cheng}
\medskip
\centerline{Institute of Physics, Academia Sinica}
\centerline{Taipei, Taiwan 115, Republic of China}
\medskip

\bigskip
\bigskip
\centerline{\bf Abstract}
\bigskip
{\small Charmed meson decays into a pseudoscalar meson $P$ and a
tensor meson $T$ are studied. The charm to tensor meson transition
form factors are evaluated in the Isgur-Scora-Grinstein-Wise
(ISGW) quark model.  It is shown that the Cabibbo-allowed decay
$D_s^+\to f_2(1270)\pi^+$ is dominated by the $W$-annihilation
contribution and has the largest branching ratio in $D\to TP$
decays. We argue that the Cabibbo-suppressed mode $D^+\to
f_2(1270)\pi^+$ should be suppressed by one order of magnitude
relative to $D_s^+\to f_2(1270)\pi^+$. When the finite width
effect of the tensor resonances is taken into account, the decay
rate of $D\to TP$ is generally enhanced by a factor of $2\sim 3$.
Except for $D_s^+\to f_2(1270)\pi^+$, the predicted branching
ratios of $D\to TP$ decays are in general too small by one to two
orders of magnitude compared to experiment. However, it is very
unlikely that the $D\to T$ transition form factors can be enhanced
by a factor of $3\sim 5$ within the ISGW quark model to account
for the discrepancy between theory and experiment. As many of the
current data are still preliminary and lack sufficient statistic
significance, more accurate measurements are needed to pin down
the issue.

}

\pagebreak

\section{Introduction}

Cabibbo-allowed and Cabibbo-suppressed two-body hadronic $D$
decays into a pseudoscalar meson $P$ and a tensor meson $T$ have
been studied in \cite{Katoch} and \cite{Munoz}, respectively. In
both studies, the charm to tensor meson transition form factors
are calculated using the ISGW (Isgur-Scora-Grinstein-Wise) quark
model \cite{ISGW}. The calculated branching ratios are of order
$10^{-5}\sim 10^{-7}$. Recently, the Cabibbo-allowed mode
$D_s^+\to f_2(1270)\pi^+$ and the Cabibbo-suppressed one $D^+\to
f_2(1270)\pi^+$ both have been measured by E791 at the level of
$10^{-3}$ \cite{E791}. More recently, FOCUS \cite{FOCUS} and BaBar
\cite{BaBar} have also reported some new measurements of $D\to TP$
decays. Though their results are still preliminary and many of
them do not have enough statistic significance (see Table I
below), the branching ratios are typically of order $10^{-3}$.
Therefore, it appears that there exists a large discrepancy
between theory and experiment. It is thus important to understand
the origin of discrepancy.

In the present work, several improvements over the previous work
\cite{Katoch,Munoz} are made. First, the charm to tensor meson
transition form factors will be calculated in the improved version
of the ISGW model \cite{ISGW2}. The updated version of this quark
model gives a more realistic description of the form-factor
momentum dependence, especially at small $q^2$. Second, the tensor
meson has a width typically of order $100-200$ MeV \cite{PDG}. The
finite width effect, which is very crucial to account for the
decays such as $D\to K_2^*(1430)\ov K$ and $D\to f_2'(1525)\ov K$
that appear to be prohibited by kinematics at first sight, is
carefully examined. Third, it is known that weak annihilation
($W$-exchange or $W$-annihilation) in charm decays can receive
sizable contributions  from nearby resonances through inelastic
final-state interactions (see e.g. \cite{a1a2Cheng}). Hence, it is
important to take into account weak annihilation contributions.

This work is organized as follows. In Sec. II we summarize the
current experimental measurements of $D\to TP$ decays. We discuss
the various physical properties of the tensor mesons in Sec. III,
for example, the decay constants and the form factors and then
analyze the $D\to TP$ decays in Sec. IV based on the generalized
factorization approach in conjunction with final-state
interactions. Conclusions are presented in Sec. V.

\section{Experimental status}
It is known that three-body decays of heavy mesons provide a rich
laboratory for studying the intermediate state resonances. The
Dalitz plot analysis is a very useful technique for this purpose.
We are interested in $D\to TP$  decays extracted from the
three-body decays of charmed mesons. Besides the earlier
measurements by ARGUS \cite{ARGUS} and E687 \cite{E687}, some
recent results are available from E791 \cite{E791}, CLEO
\cite{CLEO}, FOCUS \cite{FOCUS} and BaBar \cite{BaBar}.  The
$J^P=2^{+}$ tensor mesons that have been studied in hadronic charm
decays include $f_2(1270)$, $a_2(1320)$ and $K^*_2(1430)$. The
results of various experiments are summarized in Table I where the
product of $\B(D\to TP)$ and $\B(T\to P_1P_2)$ is shown. In order
to extract the branching ratios for the two-body decays $D\to T
P$, we need to know the branching fractions of the strong decays
of the tensor mesons \cite{PDG}:
 \be
  &&\B(f_2(1270)\to\pi\pi) =
  (84.7^{+2.4}_{-1.3})\%, \qquad~~
 \B(f_2(1270)\to K\ov K) =(4.6\pm 0.5)\%,
 \non  \\ && \B(a_2(1320)\to K\ov K)=(4.9\pm 0.8)\%, \qquad
 \B(K_2^*(1430)\to K\pi)=(49.9\pm 1.2)\%.
 \en

It is evident that most of the listed $D\to TP$ decays in Table I
have branching ratios of order $10^{-3}$, even though some of them
are Cabibbo suppressed. Note that the results from FOCUS and BaBar
are still preliminary. Indeed, many of them have not yet
sufficient statistical significance.

Note that at first sight it appears that the decay $D\to \ov
K_2^*(1430)K$ is kinematically not allowed as the $K_2^*(1430)$
mass lies outside of the phase space for the decay. Nevertheless,
it is physically allowed as $K_2^*(1430)$ has a decay width of
order $100$ MeV \cite{PDG}. Likewise, the decay $D^0\to
f_2'(1525)\ov K^0$ is also allowed.

{\squeezetable
\begin{table}[pth]
\caption{Experimental branching ratios of various $D\to TP$ decays
measured by ARGUS, E687, E791, CLEO, FOCUS and BaBar. For
simplicity and convenience, we have dropped the mass
identification for $f_2(1270)$, $a_2(1320)$ and $K^*_2(1430)$.
 }
\begin{center}
\begin{tabular}{l l l l   }
Collaboration & $\B(D\to TP)\times \B(T\to P_1P_2)$  & $\B(D\to TP)$ \\
 \hline
 E791 & $\B(D^+\to f_2\pi^+)\B(f_2\to\pi^+\pi^-)=(6.0\pm 1.1)\times 10^{-4}$ &
 $\B(D^+\to f_2\pi^+)=(1.1\pm 0.2)\times 10^{-3}$ \\
 FOCUS & $\B(D^+\to f_2\pi^+)\B(f_2\to\pi^+\pi^-)=(3.8\pm 0.8)\times 10^{-5}$ &
 $\B(D^+\to f_2\pi^+)=(6.8\pm 1.4)\times 10^{-4}$ \\
 FOCUS & $\B(D^+\to f_2\pi^+)\B(f_2\to K^+K^-)=(7.0\pm 1.9)\times 10^{-5}$ &
 $\B(D^+\to f_2\pi^+)=(3.1\pm 0.9)\times 10^{-3}$ \\
 E791 & $\B(D_s^+\to f_2\pi^+)\B(f_2\to\pi^+\pi^-)=(2.0\pm 0.7)\times 10^{-3}$ &
 $\B(D_s^+\to f_2\pi^+)=(3.5\pm 1.2)\times 10^{-3}$ \\
 FOCUS & $\B(D_s^+\to f_2\pi^+)\B(f_2\to\pi^+\pi^-)=(1.0\pm 0.3)\times 10^{-3}$ &
 $\B(D_s^+\to f_2\pi^+)=(1.8\pm 0.5)\times 10^{-3}$ \\
 ARGUS,E687 & $\B(D^0\to f_2\ov K^0)\B(f_2\to\pi^+\pi^-)=(3.2\pm 0.9)\times 10^{-3}$ &
 $\B(D^0\to f_2\ov K^0)=(4.5\pm 1.7)\times 10^{-3}$ \\
 CLEO & $\B(D^0\to f_2\ov K^0)\B(f_2\to\pi^+\pi^-)=(1.6^{+2.4}_{-1.3})\times 10^{-3}$ &
 $\B(D^0\to f_2\ov K^0)=(2.8^{+4.3}_{-2.3})\times 10^{-3}$ \\
 FOCUS & $\B(D_s^+\to f_2 K^+)\B(f_2\to\pi^+\pi^-)=(2.0\pm 1.3)\times 10^{-4}$ &
 $\B(D_s^+\to f_2 K^+)=(3.5\pm 2.3)\times 10^{-4}$ \\
 BaBar & $\B(D^0\to a_2^-\pi^+)\B(a_2^-\to K^0K^-)=(3.5\pm2.1)\times 10^{-5}$ &
 $\B(D^0\to a_2^-\pi^+)=(7.0\pm 4.3)\times 10^{-4}$ \\
 E791 & $\B(D^+\to \ov K^{*0}_2\pi^+)\B(\ov K_2^{*0}\to K^-\pi^+)=(4.6\pm 2.0)\times 10^{-4}$ &
 $\B(D^+\to \ov K^{*0}_2\pi^+)=(1.4\pm 0.6)\times 10^{-3}$ \\
 CLEO & $\B(D^0\to K^{*-}_2\pi^+)\B(K_2^{*-}\to \ov K^0\pi^-)=(6.5^{+4.2}_{-2.2})\times 10^{-4}$ &
 $\B(D^0\to K^{*-}_2\pi^+)=(2.0^{+1.3}_{-0.7})\times 10^{-3}$ \\
 BaBar & $\B(D^0\to K^{*+}_2 K^-)\B(K_2^{*+}\to K^0\pi^+)=(6.8\pm 4.2)\times 10^{-4}$ &
 $\B(D^0\to K^{*+}_2 K^-)=(2.0\pm 1.3)\times 10^{-3}$ \\
 BaBar & $\B(D^0\to \ov K^{*0}_2 K^0)\B(\ov K_2^{*0}\to K^-\pi^+)=(6.6\pm 2.7)\times 10^{-4}$ &
 $\B(D^0\to \ov K^{*0}_2 K^0)=(2.0\pm 0.8)\times 10^{-3}$ \\
\end{tabular}
\end{center}
\end{table}
}

\section{Physical Properties of scalar mesons}
The observed $J^P=2^+$ tensor mesons $f_2(1270)$, $f_2'(1525)$,
$a_2(1320)$ and $K_2^*(1430)$ form an SU(3) $1\,^3P_2$ nonet. The
$q\bar q$ content for isodoublet and isovector tensor resonances
are obvious. Just as the $\eta-\eta'$ mixing in the pseudoscalar
case, the isoscalar tensor states $f_2(1270)$ and $f'_2(1525)$
also have a mixing and their wave functions are defined by
 \be
 f_2(1270) &=& {1\over\sqrt{2}}(f_2^u+f_2^d)\cos\theta+f_2^s\sin\theta,  \non \\
 f'_2(1525) &=& {1\over\sqrt{2}}(f_2^u+f_2^d)\sin\theta-f_2^s\cos\theta,
 \en
with $f_2^q\equiv q\bar q$. Since $\pi\pi$ is the dominant decay
mode of $f_2(1270)$, whereas $f_2'(1525)$ decays predominantly
into $K\ov K$ (see Particle Data Group \cite{PDG}), it is obvious
that this mixing angle should be small. More precisely, it is
found $\theta=7.8^\circ$ \cite{PDG,Li}. Therefore, $f_2(1270)$ is
primarily an $(u\bar u+d\bar d)/\sqrt{2}$ state, while
$f'_2(1525)$ is dominantly $s\bar s$.

The polarization tensor $\vp_{\mu\nu}$ of a $^3P_2$ tensor meson
with $J^{PC}=2^{++}$ satisfies the relations
 \be
 \vp_{\mu\nu}=\vp_{\nu\mu}, \qquad \vp^{\mu}_{~\mu}=0, \qquad p_\mu
 \vp^{\mu\nu}=p_\nu\vp^{\mu\nu}=0.
 \en
Therefore,
 \be
 \la 0|(V-A)_\mu|T(\vp,p)\ra=a\vp_{\mu\nu}p^\nu+b\vp^\nu_{~\nu}
 p_\mu=0,
 \en
and hence the decay constant of the tensor meson vanishes; that
is, the tensor meson cannot be produced from the $V-A$ current.

As for the form factors, the $D\to P$ transition is defined by
\cite{BSW}
 \be \label{m.e.}
 \la P(p)|V_\mu|D(p_D)\ra = \left(p_{D\mu}+p_\mu-{m_D^2-m_{P}^2\over q^2}\,q_ \mu\right)
F_1^{DP}(q^2)+{m_D^2-m_{P}^2\over q^2}q_\mu\,F_0^{DP}(q^2),
 \en
where $q_\mu=(p_D-p)_\mu$, while the general expression for the
$D\to T$ transition has the form \cite{ISGW}
 \be \label{DTff}
 \la T(\vp,p_T)|(V-A)_\mu|D(p_D)\ra &=&
 ih(q^2)\epsilon_{\mu\nu\rho\sigma}\vp^{*\nu\alpha}p_{D\alpha}(p_D+p_T)^\rho
 (p_D-p_T)^\sigma+k(q^2)\vp^*_{\mu\nu}p_D^\nu  \non \\
 &+& b_+(q^2)\vp^*_{\alpha\beta}p_D^\alpha p_D^\beta(p_D+p_T)_\mu
 +b_-(q^2)\vp^*_{\alpha\beta}p_D^\alpha p_D^\beta(p_D-p_T)_\mu.
 \en
The form factors $k$, $b_+$ and $b_-$ can be calculated in the
ISGW quark model \cite{ISGW} and its improved version, the ISGW2
model \cite{ISGW2}. In general, the form factors evaluated in the
ISGW model are reliable only at $q^2=q^2_m\equiv (m_D-m_T)^2$, the
maximum momentum transfer. The reason is that the form-factor
$q^2$ dependence in the ISGW model is proportional to
exp[$-(q^2_m-q^2)$] and hence the form factor decreases
exponentially as a function of $(q^2_m-q^2)$. This has been
improved in the ISGW2 model in which the form factor has a more
realistic behavior at large $(q^2_m-q^2)$ which is expressed in
terms of a certain polynomial term.

The calculated $D\to T$ form factors are listed in Table II. The
form factor $h(q^2)$ is not shown there as it does not contribute
to the factorizable $D\to TP$ amplitudes. It is convenient to
express the form factors for $(D,D_s^+)\to f_2(1270)$ and
$(D,D_s^+)\to f_2'(1525)$ in terms of $D\to f_2^n$ with $n$
standing for the light non-strange quark (i.e. $D^0\to f_2^u$ for
$n=u$ and $D^+\to f_2^d$ for $n=d$) and $D_s^+\to f_2^s$
transition form factors. Note that $D\to f_2^s$ and $D_s^+\to
f_2^n$ are prohibited. In the calculations of $D\to T$ form
factors we follow \cite{Li} to use the masses: $m_{f_2^q}=1.32$
GeV and $m_{f_2^s}=1.55$ GeV.

\begin{table}[ht]
\caption{The form factors at $q^2=m_\pi^2$ calculated in the ISGW2
model, where $k$ is dimensionless and $b_+$ and $b_-$ are in units
of ${\rm GeV}^{-2}$. Shown in parentheses are the results obtained
in the ISGW model.
 }
\begin{center}
\begin{tabular}{l c c c  }
Transition & $k$ & $b_+$ & $b_-$ \\
\hline
 $D\to f_2^n$ & 0.59~(0.51) & $-0.050~(-0.083)$ & 0.061 \\
 $D_s^+\to f_2^s$ & 1.10~(1.02) & $-0.077~(-0.120)$ & 0.098 \\
 $D\to a_2(1320)$ & 0.59~(0.51) & $-0.050~(-0.083)$ & 0.061 \\
 $D\to K_2^*(1430)$ & 0.71~(0.58) & $-0.060~(-0.098)$ & 0.069  \\
\end{tabular}
\end{center}
\end{table}

Two remarks are in order. (i) The magnitude of the form factors
for the $D_s^+\to f_2^s$ transition is larger than that for $D\to
f_2^n$ owing to the larger constituent $s$ quark mass than the $u$
and $d$ quarks. That is, SU(3) symmetry breaking in $D\to f_2^n$
and $D_s^+\to f_2^s$ is sizable. (ii) The difference between ISGW
and ISGW2 model predictions for form factors at $q^2=0$ is not
significant for the charm case, though form factors in the ISGW
model fall more rapidly at small $q^2$. However, the difference
will be dramatic for the $B\to T$ case as noticed in \cite{Kim}.
For example, the $B\to a_2$ and $B\to f_2(1370)$ form factors at
$q^2=m_D^2$ obtained in the ISGW2 model are about $2-6$ times
larger than that in the ISGW model. This is because the region
covered from zero recoil to small $q^2$ in $B$ decays is much
bigger than that in $D$ decays.

\section{$D\to TP$ Decays and factorization}
We will study the $D\to TP$ decays ($T$: tensor meson, $P$:
pseudoscalar meson) within the framework of generalized
factorization in which the hadronic decay amplitude is expressed
in terms of factorizable contributions multiplied by the {\it
universal} (i.e. process independent) effective parameters $a_i$
that are renormalization scale and scheme independent. More
precisely, the weak Hamiltonian has the form
 \be
 H_{\rm eff}={G_F\over\sqrt{2}}V_{cq_1}V_{uq_2}^*\Big[ a_1(\bar uq_2)
 (\bar q_1c)+a_2(\bar q_1 q_2)(\bar uc)\Big]+h.c.,
 \en
with $(\bar q_1q_2)\equiv \bar q_1\gamma_\mu(1-\gamma_5)q_2$. For
hadronic charm decays, we shall use $a_1=1.15$ and $a_2=-0.55$\,.
Since the decay constant of tensor mesons vanishes, the
factorizable amplitude of $D\to TP$ always has the expression
 \be
 A(D\to TP) &=& i{G_F\over\sqrt{2}}V_{cq_1}V^*_{uq_2}f_P\,\vp^*_{\mu\nu}p_D^\mu p_D^\nu\,\left[
 k(m_P^2)+b_+(m_P^2)(m_D^2-m_T^2)+b_-(m_P^2)m_P^2\right] \non \\
 &\equiv& \vp^*_{\mu\nu}p_D^\mu p_D^\nu\,M(D\to TP),
 \en
where use has been made of Eq. (\ref{DTff}). The decay rate is
given by
 \be
 \Gamma(D\to TP)=\,{k_T^5\over 12\pi m_T^2}\left({m_D\over
 m_T}\right)^2|M(D\to TP)|^2,
 \en
where $k_T$ is the c.m. momentum of the tensor meson in the rest
frame of the charmed meson.

In terms of the topological amplitudes \cite{CC86}: $T$, the
color-allowed external $W$-emission tree diagram; $C$, the
color-suppressed internal $W$-emission diagram; $E$, the
$W$-exchange diagram; $A$, the $W$-annihilation diagram, the
topological quark-diagram amplitudes of various $D\to TP$ decays
are shown in Table III. There exist also penguin diagrams.
However, the penguin contributions are negligible owing to the
good approximation $V_{ud}V_{cd}^*\approx -V_{us}V_{cs}^*$ and the
smallness of $V_{ub}V_{cb}^*$. For $D\to TP$ and $D\to PT$ decays,
one can have two different external $W$-emission and internal
$W$-emission diagrams, depending on whether the emission particle
is a tensor meson or a pseudoscalar one. We thus denote the prime
amplitudes $T'$ and $C'$ for the case when the tensor meson is an
emitted particle \cite{ChengDSP}. Under the factorization
approximation, $T'=C'=0$. As pointed out in \cite{Suzuki}, the
tensor meson, for example $a_2^+$, can be produced from the tensor
operator $(\bar
u_R\gamma^\mu\stackrel{\leftrightarrow}{\partial^\nu} d_R)+(\bar
u_L\gamma^\mu\stackrel{\leftrightarrow}{\partial^\nu} d_L)$.
However, this operator must be generated by gluon corrections and
is suppressed by factors of $\alpha_s/\pi$ and $1/m_b$.

In general, $TP$ final states are suppressed relative to $PP$
states due to the less phase space available. More precisely,
 \be
 {\Gamma(D\to TP)\over \Gamma(D\to P_1P_2)}={2\over 3}\,
 {k_T^5\over k_P}\,\left({m_D\over
 m_T}\right)^4\left| {M(D\to TP)\over M(D\to P_1P_2)}\right|^2,
 \en
where $k_P$ is the c.m. momentum of the pseudoscalar meson $P_1$
or $P_2$ in the charm rest frame. The kinematic factor $h={2\over
3}\,{k_T^5\over k_P}\,\left({m_D\over m_T}\right)^4$ is typically
of order $(1-4)\times 10^{-2}$. An inspection of Table III
indicates that, in the absence of weak annihilation contributions,
the Cabibbo-allowed decays $D^+\to \ov K_2^{*0}\pi^+$ and $D^0\to
K_2^{*-}\pi^+$ will have the largest decay rates  as they proceed
through the color-allowed tree diagram $T$. It is easily seen that
all other $W$-emission amplitudes in $D\to a_2\ov K$, $D\to
f_2\pi$ and $D\to f_2 \ov K$ are suppressed for various reasons.
For example, it is suppressed by the vanishing decay constant of
the tensor meson, or by the small $f_2-f_2'$ mixing angle or by
the parameter $a_2$ or by the Cabibbo mixing angle. Let us compare
$D^+\to \ov K_2^{*0}\pi^+$ with $D^+\to \ov K^0\pi^+$
 \be
 {\Gamma(D^+\to \ov K_2^{*0}\pi^+)\over \Gamma(D^+\to \ov K^{0}\pi^+)}=
 1.3\times 10^{-2}\left({k(m_\pi^2)+b_+(m_\pi^2)(m_D^2-m_{K_2^*}^2)+b_-(m_\pi^2)m_\pi^2\over
 (m_D^2-m_K^2)F_0^{DK}(m_\pi^2)+{a_2\over
 a_1}(m_D^2-m_\pi^2)F_0^{D\pi}(m_K^2)}\right)^2.
 \en
Note that $D^+\to \ov K_2^{*0}\pi^+$ does not receive the internal
$W$-emission contribution owing to the vanishing $K_2^*$ decay
constant. The form factors $F_0^{DK}(0)$ and $F_0^{D\pi}(0)$ are
of order 0.70 \cite{BSW,MS}. Hence, the expression in the
parentheses of the above equation is of order 0.5. As a
consequence, the predicted branching ratio of $D^+\to \ov
K_2^{*0}\pi^+$ is of order $10^{-4}$, which is one order of
magnitude smaller than experiment (see Table III). As for the
decay $D^0\to K_2^{*-}\pi^+$, its branching ratio is similar to
that of $D^+\to\ov K^{*0}\pi^+$ but it receives an additional
$W$-exchange contribution. A fit of this mode to experiment will
require $|E|>|T|$, namely, $W$-exchange dominates over the
external $W$-emission, which is very unlikely. If we demand that
$|E|<|T|$, then the color-suppressed decay $D^0\to \ov
K_2^{*0}\pi^0$, which receives contributions only from the
$W$-exchange diagram, will be at most of order $10^{-5}$ (see
Table III).

For $D\to f_2(1270)\pi(K)$ decays, let us first consider $D_s^+\to
f_2\pi^+$. Its external $W$-emission amplitude is suppressed owing
to the small $s\bar s$ component in $f_2(1270)$. However,
$W$-annihilation is not subject to the $f_2-f_2'$ mixing angle
suppression. Moreover, the $D_s^+$ decay constant is much larger
than that of the pion. The magnitude of $W$-annihilation obtained
by fitting $D_s^+\to f_2\pi^+$ to the data reads
 \be \label{A/T}
 \left.{A/T}\right|_{D\to TP}\approx 0.5\, e^{-i75^\circ},
 \en
where a  relative phase of $-75^\circ$ has been assigned in analog
to $D\to PP$ [see Eq. (\ref{E/T}) below] and the tree amplitude
$T$ is referred to the one in $D_s^+\to f_2(1270)\pi^+$.

The importance of the weak annihilation contribution ($W$-exchange
or $W$-annihilation) in charm decays has been noticed long before
(see e.g. \cite{CC86,a1a2Cheng}). Even if the short-distance weak
annihilation amplitude is helicity suppressed, it does receive
long-distance contributions from nearby resonance via inelastic
final-state interactions from the leading tree or color-suppressed
amplitude. As a consequence, weak annihilation has a sizable
magnitude comparable to the color-suppressed internal $W$-emission
with a large phase relative to the tree amplitude. A quark-diagram
analysis of the Cabibbo-allowed $D\to PP$ decays yields
\cite{Rosner}
 \be \label{E/T}
 \left.A/T\right|_{D\to PP}\approx
 0.39\,e^{-i65^\circ}, \qquad\qquad
 \left.E/T\right|_{D\to PP} \approx 0.63\,e^{i115^\circ}.
 \en
We see that the ratio of $|A/T|$ in $D\to TP$ ad $D\to PP$ decays
is similar.

\begin{table}[ht]
\caption{Quark-diagram amplitudes and  branching ratios for
various $D\to TP$ decays with and without the long-distance weak
annihilation terms induced from final-state interactions.  The
$W$-annihilation amplitude $A$ is fixed by fitting to the data of
$D_s^+\to f_2(1270)\pi^+$ [see Eq. (4.6)]. The $W$-exchange
amplitude $E$ is assumed to have the expression of Eq. (4.8) for
the purpose of illustration. Experimental results are taken from
Table I and from [8]. The finite width effect of the tensor
resonances has been taken into account in theoretical
calculations.
 }
\begin{center}
\begin{tabular}{l c c c c}
Decay & Amplitude & $\B_{\rm naive}$ & $\B_{\rm FSI}$ & $\B_{\rm expt}$ \\
\hline
 $D^+\to f_2(1270)\pi^+$ & $V_{cd}V^*_{ud}(T+C+2A)\cos\theta/\sqrt{2}$
 & $2.9\times 10^{-5}$ & $2.2\times 10^{-4}$ & $(0.9\pm 0.1)\times 10^{-3}$ \\
 $D^0\to f_2(1270)\ov K^0$ & $V_{cs}V^*_{ud}(C+E)\cos\theta/\sqrt{2}$ &
 $1.0\times 10^{-4}$ & $2.5\times 10^{-4}$ & $(4.5\pm 1.7)\times 10^{-3}$ \\
 $D_s^+\to f_2(1270)\pi^+$ & $V_{cs}V_{ud}^*(T\sin\theta+2A\cos\theta/\sqrt{2})$ &
 $6.6\times 10^{-5}$ & $2.1\times 10^{-3}$ & $(2.1\pm 0.5)\times 10^{-3}$ \\
 $\quad~\to f_2(1270)K^+$ &  $V_{cs}V_{us}^*[T\sin\theta+C'\sin\theta$
 & $5.2\times 10^{-6}$ & $4.9\times 10^{-5}$
 & $(3.5\pm 2.3)\times 10^{-4}$ \\
 &  $+A(\sin\theta+\cos\theta/\sqrt{2})]$ & & & \\
 \hline
 $D^+\to f_2'(1525)\pi^+$ & $V_{cd}V^*_{ud}(T+C+2A)\sin\theta/\sqrt{2}$
 & $1.4\times 10^{-6}$ & $3.7\times 10^{-6}$ &  \\
 $D^0\to f_2'(1525)\ov K^0$ & $V_{cs}V^*_{ud}(C+E)\sin\theta/\sqrt{2}$ &
 $2.5\times 10^{-7}$ & $6.0\times 10^{-7}$ &  \\
 $D_s^+\to f_2'(1525)\pi^+$ & $V_{cs}V_{ud}^*(T\cos\theta-2A\sin\theta/\sqrt{2})$ &
 $1.6\times 10^{-4}$ & $1.5\times 10^{-4}$ &  \\
 $\quad~\to f_2'(1525)K^+$ &  $V_{cs}V_{us}^*[T\cos\theta+C'\cos\theta$
 & $4.9\times 10^{-6}$ & $7.5\times 10^{-6}$
 &  \\
 &  $+A(\cos\theta-\sin\theta/\sqrt{2})]$ & & & \\
 \hline
 $D^+\to a_2^+(1320)\ov K^0$ & $V_{cs}V_{ud}^*(T'+C)$ & $1.3\times 10^{-6}$ &
 $1.3\times 10^{-6}$ & $<3\times 10^{-3}$ \\
 $D^0\to a_2^+(1320)K^-$ & $V_{cs}V_{ud}^*(T'+E)$ & 0 & $8.9\times 10^{-8}$ & $<2\times 10^{-3}$ \\
 $\quad~\to a_2^-(1320)\pi^+$ & $V_{cd}V_{ud}^*(T+E)$ & $5.7\times 10^{-6}$ & $6.1\times 10^{-6}$ &
 $(7.0\pm 4.3)\times 10^{-4}$  \\
 \hline
 $D^+\to \ov K_2^{*0}(1430)\pi^+$ & $V_{cs}V_{ud}^*(T+C')$ & $2.6\times 10^{-4}$ &
 $2.6\times 10^{-4}$ & $(1.4\pm 0.6)\times 10^{-3}$ \\
 $D^0\to K_2^{*-}(1430)\pi^+$ & $V_{cs}V_{ud}^*(T+E)$ & $1.0\times 10^{-4}$ & $1.1\times 10^{-4}$
 & $(2.0^{+1.3}_{-0.7})\times 10^{-3}$ \\
 $\quad~\to \ov K_2^{*0}(1430)\pi^0$ & ${1\over\sqrt{2}}V_{cs}V_{ud}^*(C'+E)$ & 0 & $1.3\times 10^{-5}$
 & $<3.4\times 10^{-3}$ \\
 $\quad~\to K_2^{*+}(1430)K^-$ & $V_{cs}V_{us}^*(T'+E)$ & 0 & $1.3\times 10^{-6}$
 & $(2.0\pm 1.3)\times 10^{-3}$   \\
 $\quad~\to \ov K_2^{*0}(1430)K^0$ & $V_{cs}V_{us}^*(E_d)+V_{cd}V_{ud}^*(E_s)$
 & 0 & $\sim 10^{-8}$ & $(2.0\pm 0.8)\times 10^{-3}$
 \\
\end{tabular}
\end{center}
\end{table}

Using the $W$-annihilation term inferred from $D_s^+\to f_2\pi^+$,
we can fix the decay rates of $D^+\to f_2\pi^+$ and $D_s^+\to
f_2K^+$. Note that the predicted branching ratio for $D^+\to
f_2\pi^+$ is smaller than experiment by a factor of 4. Indeed, it
is difficult to understand why the measured branching ratio of
this mode is of the same order as $D_s^+\to f_2(1270)\pi^+$ even
the former is Cabibbo-suppressed.

$D\to f_2'(1525)\pi(K)$ decays are suppressed relative to
$f_2(1270)\pi(K)$ due to the phase space suppression. Contrary to
$D_s^+\to f_2(1270)\pi^+$, the decay $D_s^+\to f_2'(1525)\pi^+$ is
dominated by the external $W$-emission and hence it has the
largest rate among $D\to f_2'\pi(K)$ decays.

For $D\to a_2(1320)\pi(K)$ decays, both $a_2^+\ov K^0$ and
$a_2^+K^-$ are small since the factorizable external $W$-emission
vanishes owing to the vanishing $a_2$ decay constant. The decay
$D^0\to a_2^-(1320)\pi^+$ is of order $10^{-5}$ at most.

For $D\to \ov K_2^*\pi$ decays, it is found that the decay $D^+\to
\ov K_2^{*0}\pi^+$ is at most of order $10^{-4}$ as noted in
passing and it does not receive any weak annihilation
contributions. Furthermore, the unknown $W$-exchange amplitude
cannot be extracted from $D^0\to K_2^{*-}(1430)\pi^+$ or $D^0\to
f_2(1270)\ov K^0$ or $D^0\to a_2^-(1320)\pi^+$ by fitting them to
the data. It will require the unreasonable condition $|E|>|T|$.
For the purpose of illustration of the $W$-exchange effect, we
shall assume
 \be
 E/T|_{D\to TP}=0.5\,e^{i100^\circ}.
 \en

\subsection{Finite width effects}
The decay $D\to K_2^*(1430)\ov K$ is physically allowed even
though $K_2^*(1430)$ mass lies outside of the phase space for the
decay. The point is that $K_2^*(1430)$ has a decay width of order
$100$ MeV \cite{PDG} and hence it is necessary to take into
account the finite width effect. Likewise, the decay $D^0\to
f_2'(1525)\ov K^0$ which is outside of phase space also can occur.

The measured decay widths of various tensor mesons are given by
\cite{PDG}
 \be
 && \Gamma_{f_2(1270)}=185.1^{+3.4}_{-2.6}\,{\rm MeV}, \quad
 \Gamma_{f_2'(1525)}=76\pm 10\,{\rm MeV}, \quad
 \Gamma_{a_2(1320)}=107\pm 5\,{\rm MeV}, \non \\
 && \Gamma_{K_2^{*\pm}(1430)}=98.5\pm2.7 \,{\rm MeV}, \quad
 \Gamma_{K_2^{*0}(1430)}=109\pm 5 \,{\rm MeV}.
 \en
To take into account the finite width effect of the tensor
resonances, we employ the factorization relation to ``define" the
$D\to TP$ decay rate
 \be \label{fact}
 \Gamma(D\to TP\to P_1P_2P)=\Gamma(D\to TP)\B(T\to P_1P_2),
 \en
with
 \be \label{3body}
 \Gamma(D\to TP\to P_1P_2P) &=& {1\over
 2m_D}\int^{(m_D-m_P)^2}_{(m_1+m_2)^2}{dq^2\over 2\pi}\,|\la
 TP|{\cal H}_W|D\ra|^2\,{\lambda^{1/2}(m_D^2,q^2,m_P^2)\over 8\pi m_D^2}  \non \\
 &\times& {1\over (q^2-m_T^2)^2+(\Gamma_{12}(q^2)m_T)^2}\,g^2_{TP_1P_2}
 {\lambda^{1/2}(q^2,m_1^2,m_2^2)\over 8\pi q^2},
 \en
where $\lambda$ is the usual triangluar function
$\lambda(a,b,c)=a^2+b^2+c^2-2ab-2ac-2bc$, $m_1$ ($m_2$) is the
mass of $P_1$ ($P_2$), $g_{TP_1P_2}$ is the strong coupling to be
defined below, and the ``running" or ``comoving" width
$\Gamma_{12}(q^2)$ is a function of the invariant mass
$m_{12}=\sqrt{q^2}$ of the $P_1P_2$ system and it has the
expression \cite{Pilkuhn}
 \be
 \Gamma_{12}(q^2)=\Gamma_T\,{m_T\over m_{12}}\left({p'(q^2)\over
 p'(m_T^2)}\right)^5\,{9+3R^2p'^2(m_T^2)+R^4p'^4(m_T^2)\over 9+
 3R^2p'^2(q^2)+R^4p'^4(q^2)},
 \en
with $p'(q^2) = \lambda^{1/2}(q^2,m_1^2,m_2^2)/(2\sqrt{q^2})$. We
shall follow \cite{CLEO} to take $R$, the ``radius" of the meson,
to be $1.5\,{\rm GeV}^{-1}$. From the measured decay width of the
tensor meson, one can determine the strong coupling $g_{TP_1P_2}$
via
 \be
 \Gamma(T\to P_1P_2)={g_{TP_1P_2}^2 m_T\over 15\pi}\,\left({p_c\over
 m_T}\right)^5,
 \en
where $p_c$ is the c.m. momentum of $P_1$ and $P_2$ in the rest
frame of the tensor meson.

Note that in the narrow width approximation, one can show that the
factorization relation (\ref{fact}) holds. When the decay width is
not negligible we will use Eq. (\ref{3body}) to evaluate the
three-body decay $\Gamma(D\to TP\to P_1P_2P)$ and employ Eq.
(\ref{fact}) to define the decay rate of $D\to TP$. To evaluate
the decay rate of $D\to TP\to P_1P_2P$, we will assume that
$g_{TP_1P_2}$ is insensitive to the $q^2$ dependence when the
resonance is off its mass shell. Numerically it is found that when
the finite decay width of the tensor meson is taken into account,
the decay rate of $D\to TP$ is generally enhanced by a factor of
$2\sim 3$. The results of the calculated branching ratios shown in
Table III have included finite width effects.

\section{Discussion and Conclusion}
Charmed meson decays into a pseudoscalar meson and a tensor meson
are studied. The charm to tensor meson transition form factors are
evaluated in the Isgur-Scora-Grinstein-Wise quark model. The main
conclusions are:
 \begin{itemize}
 \item
The external $W$-emission contribution to the decay $D_s^+\to
f_2(1270)\pi^+$ is suppressed by the fact that $f_2(1270)$ is
predominately $n\bar n$. Hence, this decay is dominated by the
$W$-annihilation contribution. We argue that the
Cabibbo-suppressed mode $D^+\to f_2\pi^+$ should be suppressed by
one order of magnitude relative to $D_s^+\to f_2(1270)\pi^+$,
contrary to the E791 measured results.
 \item
The long-distance $W$-annihilation contributions induced from
nearby resonances via inelastic final-state interactions gives the
dominant contributions to $(D^+,D_s^+)\to f_2(1270)\pi^+$,
$D_s^+\to f_2(1270)K^+$. Under the factorization approximation,
the decays $D^0\to a_2^+(1320)K^-,~\ov
K_2^{*0}(1430)\pi^0,~K_2^{*+}(1430)K^-$ receive contributions
solely from the $W$-exchange diagram.
  \item
Among the $D\to TP$ decays, $D_s^+\to f_2(1270)\pi^+$ has the
largest branching ratio of order $10^{-3}$. The modes $D^+\to
f_2(1270)\pi^+$, $D^0\to f_2(1270)\ov K^0$, $D_s^+\to
f_2'(1525)\pi^+$, $D^+\to \ov K^{*0}\pi^+$ and $D^0\to
K_2^{*-}\pi^+$ are of order $10^{-4}$.
  \item
  The decay rate of $D\to TP$ is generally enhanced by a factor of
$2\sim 3$ when the finite width effect of the tensor resonances is
taken into account. In particular, it is necessary to include the
finite width effect to explain the decays $D\to K_2^*(1430)\ov K$
and $D\to f_2'(1525)K$.
  \item
Except for the Cabibbo-allowed decay $D_s^+\to f_2(1270)\pi^+$,
the predicted branching ratios of $D\to TP$ decays are in general
too small by one to two orders of magnitude compared to
experiment.  However, it is very unlikely that one can enhance the
$D\to T$ transition form factors within the ISGW quark model by a
factor of $3\sim 5$ to account for the discrepancy between theory
and experiment. As many of the current data have not yet enough
statistical significance, it is important to have more accurate
measurements in the near future to pin down the issue.

  \end{itemize}

\vskip 2.5cm \acknowledgments This work was supported in part by
the National Science Council of R.O.C. under Grant No.
NSC91-2112-M-001-038.


\end{document}